# Circuits and excitations to enable Brownian token-based computing with skyrmions

Maarten A. Brems[1], Mathias Kläui[1,*] and Peter Virnau[1]

[1] *Institute of Physics, Johannes Gutenberg-Universität Mainz, 55099 Mainz, Germany*

*\* Electronic mail: klaeui@uni-mainz.de*

## Abstract

Brownian computing exploits thermal motion of discrete signal carriers (tokens) for computations. In this paper we address two major challenges that hinder competitive realizations of circuits and application of Brownian token-based computing in actual devices for instance based on magnetic skyrmions. To overcome the problem that crossings generate for the fabrication of circuits, we design a crossing-free layout for a composite half-adder module. This layout greatly simplifies experimental implementations as wire crossings are effectively avoided. Additionally, our design is shorter to speed up computations compared to conventional designs. To address the key issue of slow computation based on thermal excitations, we propose to overlay artificial diffusion induced by an external excitation mechanism. For instance, if magnetic skyrmions are used as tokens, artificially induced diffusion by spin-orbit torques or other mechanisms increases the speed of computations by several orders of magnitude. Combined with conventional Brownian computing the latter could greatly enhance the application scenarios of token-based computing for instance for low power devices such as autonomous sensors with limited power that is harvested from the environment.

Brownian computing is a method for logic operations, inspired by noise-exploiting mechanisms in biological processes, e.g., as observed in molecular motors[1–4]. Exploitation of thermal noise for parts of computations offers great application potential for sensors, which can harvest thermal energy from the environment. Thermal fluctuations also increasingly pose challenges in the miniaturization of computing devices[5–7], and Brownian computing may at least conceptually turn this problem into an advantage.

Magnetic skyrmions are particularly promising token candidates[8–10], which act as stable and discrete signal carriers in Brownian computing[11–14]. These two-dimensional, topologically stabilized whirls of magnetization not only exhibit quasi-particle behavior[15–20], but undergo thermally activated diffusion[8,9,21,22]. They have been stabilized in magnetic films and bulk materials at temperatures from a few Kelvin to far above room temperature[15,16,23–28] even without application of an external field[29,30], and recent proposals for device applications include skyrmion-based logic[21,31,32] and racetrack memory[15,33,34]. Skyrmions can also be manipulated by a multitude of mechanisms including external fields as well as their gradients[35–39], and spin torques[40–46], which is also a desirable feature in this context.

In this manuscript we address major challenges of Brownian token-based computing that occur for many implementations, and we provide details and estimates of the performance improvement for the concrete realization using magnetic skyrmions. Conventional circuit layouts (as discussed below) include crossings of wires which pose an extraordinary implementation obstacle when two-dimensional or quasi two-dimensional tokens are employed. Here, we present an exemplary layout of a half-adder, which completely does without wire crossings. An obvious challenge in the application of Brownian computing is the inherent non-deterministic computation time, which typically exceeds those of conventional approaches by orders of magnitude. To address this challenge, we propose to overlay or replace the natural diffusion of the signal carrier with an artificial diffusion, induced by an external excitation mechanism that stimulates random token movement. As the latter is in principle only limited by the frequency and amplitude of the external stimulus enormous speed-ups can be realized at the cost of an increased energy consumption used for the driving mechanism. Induced and natural diffusion modes could of course also be used side by side in corresponding applications. In the following considerations we will employ magnetic skyrmions as tokens when evaluating the concrete performance improvements even though our conclusions are in principle universal and independent of the nature of the signal carrier.

In Brownian token-based computing, a computation is performed as discrete and indivisible signal carriers (tokens) traverse a circuit, which connects a set of input lines to a set of output lines (Fig. 1)[11]. The circuit can be understood as a network of interconnects and modules with special transition rules, which ensure that every input leads to the correct output. Input and output signals are set and read by dual rail encoding[47]. So there exist a 1-line and a 0-line for each signal and the presence of a token on one (and therefore absence on the other) determines the signal value (Fig. 1). The fluctuating tokens perform Brownian random search for the circuit paths which advance the computation[12,48]. The circuits' transition rules are manifested in the primitive circuit modules. We employ the set of 3 primitive modules proposed in Ref.[12] generating circuits which conserve the number of tokens and are robust against delays[49]. The hub (circle in Fig. 1) is a junction of wires, which allows tokens to transition from one wire to another. The ratchet (line with triangle) is an optional module like a wire, used to speed up the computation by favoring one direction of motion. As the circuits are asynchronous one must substitute for a clock by a synchronization module[50]. The cjoin (square) only allows for a token passage in pairs of two, where the tokens pass from the two occupied lines to the two free lines. We employ an implementation of the cjoin in which a token that arrives first is locked inside the cjoin for a fixed lock-up time. If a second token arrives at the cjoin during this time, both tokens pass the cjoin and else the locked-up token is released to its original path. A discussion of recently proposed potential implementations of the necessary modules for skyrmion tokens can be found in the supplementary material.

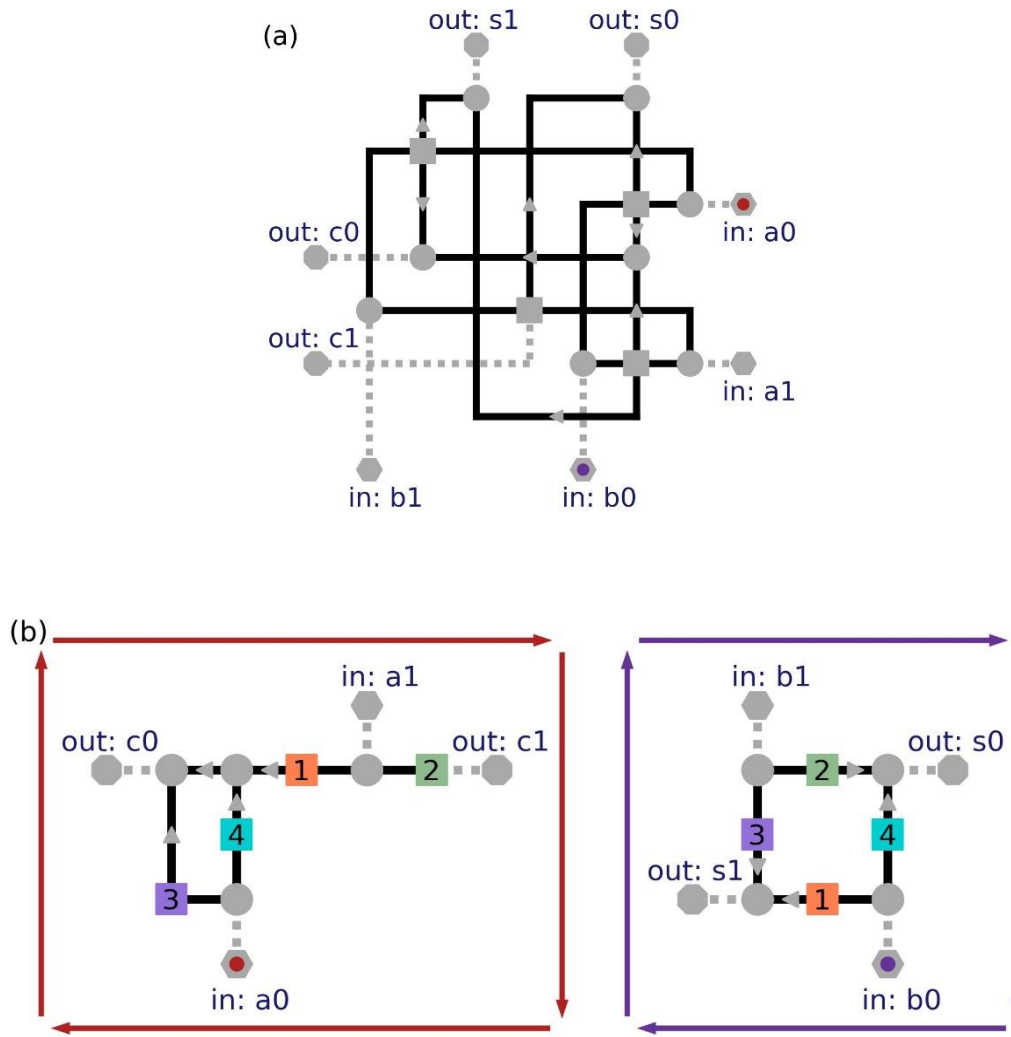

FIG. 1: Two layouts for a half-adder containing hubs (grey circles) and ratchets (lines with triangles, the triangle tip indicates the preferred direction of motion). The set input is 0+0 in both cases. (a) Layout with conventional cjoins (squares) containing wire-crossings. Reproduced from T. Nozaki, Y. Jibiki, M. Goto, E. Tamura, T. Nozaki, H. Kubota, A. Fukushima, S. Yuasa, and Y. Suzuki, Appl. Phys. Lett. **114**, 012402 (2019), with the permission of AIP Publishing. (b) Crossing-free layout in which cjoin-halves (colored squares) with same color/number can only be passed together. The colored arrows show the relevant directions from which the direction of the induced movement is chosen randomly. Since the circuit is separated into two subcircuits, each token can be moved independently, indicated by the set of arrows with the same color as the tokens (bold dots).

The half-adder circuit has been studied for two distinct reasons. First, it is a high-relevance composite module as it is used to add two bits. It can be used in many more complex Brownian circuits like conditional counters and counting memories[12]. Second, the half-adder is one of the simplest Brownian circuits which still exhibits all fundamental features of Brownian computing like a plurality of computational paths and possible dead-lock situations for non-fluctuating tokens. Thus, it is an illustrative example circuit for the method of token-based computing exploiting artificially induced diffusion. A half-adder circuit layout reproduced from Ref.[8] is depicted in Fig. 1(a). The interconnection of the primitive modules to form a half-adder has been proposed in Ref.[12]. The dual rail encoded output bits "s" and "c" represent the sum ($2^0$) and carry ($2^1$) digit of the computation result, respectively. For effectively two-dimensional skyrmion systems, crossings in the circuit layout as they occur in Fig. 1(a) pose a severe challenge for experimental implementation. Therefore, we propose a crossing-free layout (Fig.1(b)) for the half-adder obtained by altering the implementation of the cjoin module. To obtain this layout the cjoin is separated into two cjoin-halves (colored squares labeled with integers). These halves must still be communicating to that in each cjoin-half the token can only pass if another token also passes the other half. So, synchronization of the cjoin halves must be ensured over longer distances. This can be done by electronics integrated with the sample as an active control is necessary for the operation of a cjoin in any case (even for the conventional design). For the non-separated cjoin, the interactions between skyrmions can make the reliable operation of the cjoin challenging. Separating their trajectories physically is also an option to deal with this issue.

We use a minimalistic random walk-based[51,52] simulation algorithm that has been developed to estimate the computation time distribution and thereby the mean computation time for the crossing-free half-adder layout. We employ a one-dimensional random walk model in which the step size in space $\delta x$ is chosen to be the length of the shortest wire and $\delta t$ is the required time to perform one step. The size of $\delta x$ in experimental units is thus determined by the circuits' extend. The time step $\delta t$ can be obtained from the temperature dependent skyrmion diffusion coefficient. One way to obtain the relation is by matching the scaling of the mean squared displacement of the walker $\langle [\Delta x(\Delta t)]^2 \rangle =$ $(\delta x^2/\delta t) \cdot \Delta t$ to the one skyrmion diffusion $\langle[\Delta x(\Delta t)]^2\rangle = 2D \cdot \Delta t$. Here, $[\Delta x(\Delta t)]^2$ is squared spacial displacement during the time $\Delta t$, $D$ is the diffusion coefficient and angled brackets indicate the average. The computation time is then translated to experimental units via $\delta t = \delta x^2/(2D)$. A hub is modelled to allow for token movement into each attached wire with equal probability. A first token entering a cjoin will be locked inside for a fixed lock-up time as discussed above. For all presented simulations, this lock-up time has been chosen to be the integer multiple of timesteps $\delta t$ resulting in the minimum mean computation time. Ratchets are not considered in the model for better comparability to the results for the induced diffusion method. Instead, part of the ratchets' functionality is incorporated by making the cjoins unidirectional. However, a natural extension for the model would be biased random walks in the ratchets. The resulting computation time distribution for input 0+0 is shown in Fig. 2. In Brownian computing the computation time generally depends on the input due to the differences in accessible path lengths. Here we only discuss the input 0+0 as the distributions for the other inputs are of similar shape and the mean computation times of the same order of magnitude. The distribution (Fig. 2) is strongly peaked, but also has a slow decaying tail which shifts the mean computation time away from the peak. This illustrates a severe disadvantage of Brownian computing for many applications on time scales of the order of the mean computation time: computation times larger than twice the mean computation time still occur frequently. One way to circumvent this issue is to speed up computations on demand as discussed below. Rough estimates for the mean computation time translated to experimental units are given in Table I for different temperatures and shortest wire lengths. The temperature dependence of the diffusion coefficient has been taken from previous measurements in a low pinning Ta/CoFeB/Ta/MgO/Ta multilayer system[21]. In this estimate, the mean computation time can range from a few minutes to many hours depending on the temperature. This drastic effect is due to the exponential temperature dependence of the skyrmion diffusion coefficient. Additionally, shorter wire lengths can significantly speed up the computation, which is another advantage of the crossing-free layout.

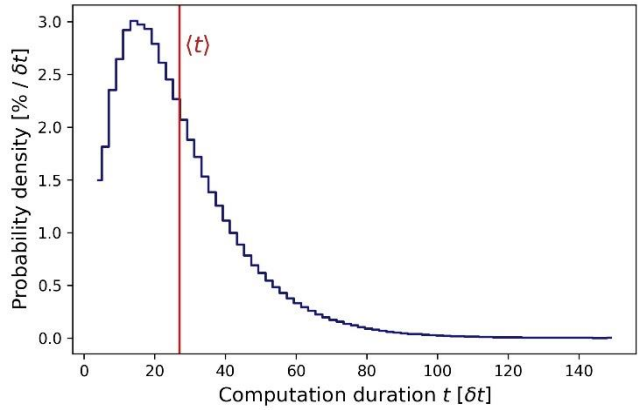

FIG. 2: Normalized computation time distribution of $3 \times 10^6$ simulated single computations of a half-adder exploiting thermal diffusion. The layout from Fig. 1(b) is chosen and the input is 0+0. A red vertical line indicates the mean computation time.

Results so far indicate that Brownian computing using skyrmions can be rather slow, which is expected from a highly energy-efficient method relying predominantly on thermal fluctuation for the computation. While certainly various applications exist without the need for high performance, many of those are still time-critical in the sense that a computation is required once per certain time interval. Exemplary, an autonomous sensor is discussed below. If the time interval is of the order of the mean computation time, the applicability of skyrmion-based Brownian computing in such scenarios is heavily reduced due to non-deterministic computation time. Thus, a method to speed up Brownian computing as necessary or in general is required. We propose a method of token-based computing which exploits artificially induced diffusion instead of or in addition to thermal diffusion. The tokens are moved by means of induced dynamics, e.g., by an external mechanism. Therein the direction of induced motion is chosen at random from the set of relevant directions for the circuit (colored arrows in Fig.1(b)). The set of relevant directions must be chosen in such a way that the system is ergodic i.e., every position on the computational paths can be reached. This is necessary to provide the backtracking ability. Induced dynamics can be applied to all tokens in the circuit or only in some paths. For the crossing-free half-adder (Fig. 1(b)) both tokens can be moved independently even using the former approach due to the separated subcircuits. The method also works for correlated token movement, but one must usually increase the cjoin lock-up time. In principle, this method can be applied to any system in which token dynamics can be induced by means of an external mechanism. Magnetic skyrmions are promising candidates for induced diffusion in token-based computing as there exists a multitude of different mechanisms for induced dynamics. Using different stimulus mechanisms, suitable methods include e.g. motion due to gradients of magnetic fields[35–37] and motion due to electric fields, e.g., by means of electric field induced anisotropy variations in the system[38,39]. Effective acting spin-polarized currents can induce spin torques like spin transfer torques[43,44] or spin orbit torques[45,46] to efficiently move skyrmions. More mechanisms for induced skyrmion dynamics can be found in Ref.[15]. At the price of external energy, such mechanisms can induce directed motion at much higher speeds than what has been measured for skyrmion diffusion[21,42,53]. While it has recently been proposed to employ skyrmion diffusion enhanced by an alternating magnetic field[14], here the induced motion itself is deterministic but its direction is changed randomly at certain times. Thereby, the length scales on which motion is deterministic and on which it is random are controllable. The speed of the computation is mainly limited by the speed of the induced motion and the condition not to destroy the tokens in the process. Skyrmion velocities above 100 m/s have been measured in low pinning multilayer systems[42,53]. However, the induced motion will often be directed towards the boundary of the confinement of, e.g., a wire or a cjoin. Therefore, the peak skyrmion velocity should not be chosen higher than the upper critical annihilation velocity at the boundary. To roughly estimate the magnitude of this annihilation velocity $v_c$ we employ a result from a numerical study of Iwasaki et al.[54] $v_c \approx 10^3 (D/J)^2 \ m/s$. Here $D$ is the Dzylashinskii-Moriya interaction parameter, $J$ is the exchange coupling strength and $m/s$ are meters per second. For typical parameters from Refs.[55,56] e.g., an exchange stiffness $A_s = 10 \ pJ/m$ for CoFeB at a thickness of $t = 1 \ nm$ with $J \approx A_s/t$ for a multilayer stack and $D = 1.5 \ mJ/m^2$ one obtains $v_c \approx 22 \ m/s$. As in many mechanisms the induced peak velocity differs from the average velocity, we list in the mean computation time overview (Table I) also values for an average velocity of $\langle v \rangle \approx 6.4 \ m/s$. This corresponds to a $13 \ ns$ Gaussian velocity profile due to a current as in the supplementary material of Ref.[53], scaled for the peak velocity to match the annihilation velocity $v_c$.

To obtain computation time estimates we modify the previous simulation model to account for induced diffusion instead of thermal diffusion. Here, the tokens do not move if the direction of induced movement is perpendicular to a confining boundary. Ratchets are not employed as the induced motion will usually dominate over the ratchet motion bias. Note that the same model is now less minimalistic, as the system can and should always be tuned such that the tokens are moved over the entire length of the shortest wire $\delta x$ by one application of the movement inducing mechanism. (E.g., one current pulse.) This choice is made as a random search inside a wire is not required for successful computation, contrary to the randomness of motion on the scale of the interconnection between different cjoins. Thus, the spacial step size of the random walk is still $\delta x$, but the time required for one step is now $\delta t_i = \delta x / \langle v \rangle$ instead of $\delta t_t = \delta x^2 / (2D)$. The shape of the computation time distribution is very similar to the one for thermal diffusion depicted in Fig. 2. The mean computation time is however about 5 orders of magnitude smaller for the induced diffusion method such that the times are now in the order of tens of microseconds instead of minutes (Table I). Additionally, the mean computation time scales only linearly with the wire length for the induced diffusion method instead of quadratic as for conventional Brownian computing. Thus, the induced diffusion method is even relatively faster for larger circuits. An approximate estimate of the mean electric energy per computation required for induced diffusion can be found in the supplementary material. Another advantage is that the duration, strength and direction choosing probability of the external movement inducing excitation must neither be identical for all relevant directions of induced movement, nor must they be strictly consistent over time. Conversely, one can adapt the excitation to the circuit geometry. E.g., if the circuits' extend in y-direction is much larger than in x-direction, one can make random induced movement more frequent in the former direction than in the latter. A feature which would require anisotropic diffusion methods for conventional Brownian computing[22].

Moreover, induced diffusion is independent of thermal diffusion. It can thus be applied in addition to thermal diffusion or replacing it. Brownian computing can thereby be achieved even in systems and materials without significant natural diffusion. Also, the external excitation can be tuned such that the circuit matches a given power consumption. Most importantly, induced diffusion can be activated as necessary for a conventional Brownian circuit. As an example, we consider an autonomous sensor that should provide a measured value once per fixed time interval. Autonomous sensors are usually powered by a small energy harvesting unit for generating electric energy from the environment. Therefore, the sensor must be very energy efficient. This is sometimes considered to be a probable application for conventional Brownian computing, given the time interval is in the order of the mean computation time of a conventional Brownian circuit. Yet sometimes the sensor will have not yet provided a result by the end of the interval due to the non-deterministic computation time, which still renders conventional Brownian computing useless for this task. This issue can be circumvented by activating artificially induced diffusion as necessary to ensure that no interval is skipped. The variation in the induced diffusion computation time is irrelevant on this time scale as the induced diffusion method is so much faster than conventional Brownian computing. This way, the circuit predominantly functions based on thermal diffusion. Only when there is the need to speed-up the computation, induced diffusion is activated and some of the energy reserves of the harvesting unit are used.

| Method | Condition | $\delta x$ = 10 µm | $\delta x$ = 5 µm |
|---|---|---|---|
| Thermal Diffusion | T = 287.6 K | 18.7 h | 4.6 h |
| Thermal Diffusion | T = 307.6 K | 4 min | 1 min |
| Induced Diffusion | $\langle v \rangle \approx 6.4\ m/s$ | 133.6 µs | 66.8 µs |
| Induced Diffusion | $\langle v \rangle \approx 22\ m/s$ | 38.9 µs | 19.4 µs |

TABLE I: Overview of estimated mean computation times for input 0+0 in experimental units depending on the shortest wire length $\delta x$ for thermal and induced diffusion. The average is taken over $3 \times 10^6$ simulated single computations. For thermal diffusion, the mean computation time depends

on the temperature dependent diffusion coefficient (Ref.[21]). For induced diffusion, the mean computation time depends on the average skyrmion velocity.

In conclusion, we address two key challenges for implementing Brownian token-based computing and assess the performance gains entailed. We present a crossing-free layout for a skyrmion-based Brownian half-adder providing greatly simplified experimental implementation and faster computations. We illustrate the non-deterministic computation time in Brownian computing by minimalistic random walk-based simulations and provide estimates for the mean computation time of the half-adder. Moreover, we present a method of token-based computing using skyrmions. Here, artificial diffusion is induced by an external mechanism such as electric or magnetic fields, their gradients or spin torques. Thereby, the computation time is decreased by several orders of magnitude at the expense of external energy. This method allows to accelerate conventional Brownian computing as necessary, match the computation speed to a given energy consumption and allow for Brownian computing even in systems without significant thermal fluctuations. In this way, the range of applications for which Brownian computing becomes viable, is significantly extended.

See supplementary material for a discussion of potential implementations of the necessary modules for skyrmion tokens and an estimate of the mean electric energy per computation required for induced diffusion.

We are grateful to the Deutsche Forschungsgemeinschaft (DFG, German Research Foundation) for funding this research: Project number 233630050-TRR 146 and Project number 403502522-SPP 2137 Skyrmionics. The authors gratefully acknowledge computing time granted on the HPC cluster Mogon at Johannes Gutenberg University Mainz. The authors furthermore acknowledge funding from TopDyn,

SFB TRR 173 Spin+X (project A01 #268565370), and from the Horizon 2020 framework program of the European commission under grant No. 856538 (ERC-SyG 3D MAGIC).

The data that support the findings of this study are available from the corresponding author upon reasonable request.